\documentclass[a4paper,fleqn]{cas-sc}

\usepackage[authoryear,longnamesfirst]{natbib}

\usepackage{lineno}

\def\tsc#1{\csdef{#1}{\textsc{\lowercase{#1}}\xspace}}
\tsc{WGM}
\tsc{QE}
\begin{document}
\let\WriteBookmarks\relax
\def\floatpagepagefraction{1}
\def\textpagefraction{.001}

\shorttitle{}    

\shortauthors{}  

\title [mode = title]{ The Sensitivity to initial conditions of the Co-orbital outcomes of Lunar Ejecta}  




\author[1]{Jose Daniel Castro-Cisneros}[
    orcid=0000-0002-6624-4214
]

\cormark[1]


\ead{jdcastrocisneros@arizona.edu}


\credit{Conceptualization, Data curation, Formal analysis, Investigation, Software, Visualization, Writing - original draft, Writing - review \& editing}

\affiliation[1]{organization={Physics Department, The University of Arizona},%Department and Organization
            addressline={1118 E. Fourth Street}, 
            city={Tucson},
            postcode={85721}, 
            state={AZ},
            country={USA}}

\author[2]{Renu Malhotra}

% Footnote of the second author

% Email id of the second author
%\ead{}


\credit{Conceptualization,Funding acquisition, Supervision, Writing - original draft, Writing - review \& editing}

\affiliation[inst2]{organization={Lunar and Planetary Laboratory, The University of Arizona},%Department and Organization
            addressline={1629 E. University Blvd.}, 
            city={Tucson},
            postcode={85721}, 
            state={AZ},
            country={USA}}


\author[3]{Aaron J. Rosengren}
\credit{Supervision, Writing - original draft, Writing - review \& editing}

\affiliation[inst3]{organization={Mechanical and Space Engineering, UC San Diego},%Department and Organization
            addressline={9500 Gilman Drive}, 
            city={La Jolla},
            postcode={92093}, 
            state={CA},
            country={USA}}

\cortext[cor1]{Corresponding author}


\begin{abstract}
Lunar ejecta, produced by meteoroidal impacts, have been proposed for the origin of the near-Earth asteroid (469219) Kamo’oalewa, supported by its unusually Earth-like orbit and L-type reflectance spectrum (Sharkey et al., 2021). 
In a recent study (Castro-Cisneros et al. 2023), we found with N-body numerical simulations that the orbit of Kamo’oalewa is dynamically compatible with rare pathways of lunar ejecta captured into Earth's co-orbital region, persistently transitioning between horseshoe and quasi-satellite (HS-QS) states. 
Subsequently, Jiao et al. (2024) found with hydrodynamic and N-body simulations that the geologically young lunar crater Giordano Bruno generated up to 300 Kamo’oalewa-sized escaping fragments, and up to three of those could have become Earth co-orbitals.
However, these results are based upon specific initial conditions of the major planets in the Solar System, close to the current epoch.
In particular, over megayear time spans, Earth’s eccentricity undergoes excursions up to five times its current value, potentially affecting the chaotic orbital evolution of lunar ejecta and their capture into Earth's co-orbital regions. 
In the present work, we carry out additional numerical simulations to compute the statistics of co-orbital outcomes across different launch epochs, representative of the full range of Earth’s eccentricity values.
Our main results are as follows:
Kamo'oalewa-like co-orbital outcomes of lunar ejecta vary only slightly across the range of Earth's orbital eccentricity, suggesting no privileged ejecta launching epoch for such objects;
the probability of co-orbital outcomes decreases rapidly with increasing launch speed, but long-lived HS-QS states are favored at higher launch speeds.  \nocite{*}%% Remove this line from your manuscript.
\end{abstract}


\begin{highlights}
\item Co-orbital outcomes sharply decline as lunar ejecta launch speed increases. 
\item High speed ejecta favor long-lived quasi-satellite co-orbital outcomes. 
%\item Lunar ejecta may reach co-orbital states across Earth's full eccentricity range.
\item Co-orbiter production by lunar ejecta is largely unaffected by Earth's eccentricity at the time of launch.

\end{highlights}

% Keywords

\maketitle

%\begin{linenumbers}

% Main text
\section{Introduction}
\label{sec:Introduction}

\begin{figure}
    \centering
    \includegraphics[scale=1.0]{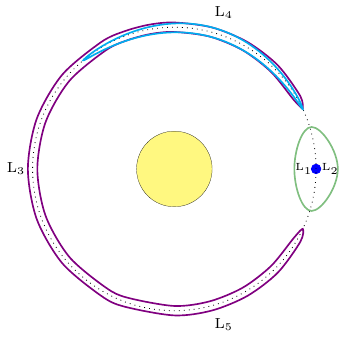}
    \caption{ The three classes of co-orbitals considered in this work: Trojans/tadpoles (\textit{cyan}) oscillate about the $L_{4}$ or $L_{5}$ Lagrange points; horseshoe (\textit{violet}) companions oscillate about the $L_3$ Lagrange point, diametrically opposite the planet's location, and encompass both $L_4$ and $L_5$ Lagrange points; and quasi-satellites (\textit{green}) orbit outside the planet's Hill sphere and enclose both the collinear $L_1$ and $L_2$ Lagrange points. 
    }
    \label{fig:orbits}
\end{figure}

Small bodies may share the orbit of a planet in a long-term stable configuration by librating in the 1:1 mean-motion resonance \citep{cMsD99}; such configurations are referred to as co-orbital motion.  In the context of the idealized, circular, restricted three-body problem (CR3BP), there are three main types of co-orbital states: Trojan/tadpole (T), horseshoe (HS), and retrograde satellite/quasi-satellite (QS) \citep{fNetal99}. These cases are portrayed in Fig.~\ref{fig:orbits}, and can be characterized by oscillations in the mean longitude  of the object relative to the secondary body (host planet). Trojan/tadpole co-orbitals oscillate around a relative mean longitude $\Delta \lambda = \lambda - \lambda_{\mathrm{host}}$  of $60^{\circ}$ or $300^{\circ}$; HS co-orbitals oscillate around $180^{\circ}$ and QS co-orbitals do it around $0^{\circ}$.

The near-Earth object (469219) Kamo'oalewa is an exceptionally long-lived co-orbital of the Earth, presently in a QS state. Numerical computations indicate that persistently transitions between QS and HS states for millions of years \citep{bD19,cFM16a}
%Several hypotheses for Kamo'oalewa's origin have been proposed based on the available information of its physical properties. In particular, 
Sharkey and colleagues measured its reflectance spectrum and found it to have a L-type profile, which resembles lunar silicates \citep{bS21}. This poses the novel possibility that Kamo'oalewa may be lunar ejecta from a meteoroidal impact. 

This hypothesis has been tested in \citet{Castro2023} by numerically simulating test particles launched from the Moon's surface and following their subsequent orbital evolution under the influence of the eight major planets in the Solar System and the Moon. In contrast with previous investigations of lunar ejecta conducted by other authors \citep[e.g.,][]{bG95}, which focused on detecting whether lunar ejecta collide with the Earth and Moon or escape into heliocentric orbits, \citet{Castro2023} look for detecting the transfer of lunar ejects into co-orbital states. This requires higher accuracy and high time resolution N-body simulations. The results show that a small fraction of the launched particles can evolve into co-orbital bodies exhibiting transitions between HS and QS state for time spans of at least tens of thousands of years. The calculations presented are however limited in the sense that they are valid for initial conditions of the Solar System near to the current epoch. A larger exploration needs to be done to identify a sensitivity to the initial epoch. 

Possibly the most significant factor influencing these co-orbital outcomes of lunar ejecta is the secular variation of Earth's eccentricity. Over megayear time spans, Earth's eccentricity undergoes excursions up to five times its current value, making its orbital velocity to vary roughly by $2 \, \mathrm{km \, s^{-1}}$ (compared with only about $0.5 \, \mathrm{km \, s^{-1}}$ at the current epoch). This is a significant fraction of the launch speeds of escaping lunar ejecta which are typically just a few $\mathrm{km \, s^{-1}}$. Therefore in this work we have investigated the sensitivity of the coorbital outcomes of lunar ejecta to the Earth's eccentricity at the time of ejecta launch. 
The rest of the paper is organized as follows: In \textsection\ref{sec:Methodology} we
explain how to extend the methods used in \citep{Castro2023} and our criteria for the identification of co-orbitals. The results of the numerical experiments are described in \textsection\ref{sec:results}. Their analyses and a discussion of how they help to understand the origin of objects like Kamo'oalewa is given in \textsection\ref{sec:Discussion}.

\section{Methodology}
\label{sec:Methodology}

\subsection{Numerical model}
\label{subsec:model}

We employ an extension of the methods previously used by \cite{Castro2023} to explore the dynamical fates of lunar ejecta, akin to the techniques adopted by \cite{aD10} for satellite ejecta within the Saturnian system. 
Our model considers the gravitational interactions of the eight major planets in the Solar System, and the Sun, and the Moon. Test Particles (TP's) are launched from the surface of the Moon and propagated using the \texttt{IAS15} integrator within \texttt{REBOUND}. The \texttt{DIRECT} predefined module was used to detect collisions with the massive bodies. An initial step size of 1.2 days was used and the accuracy parameter was set to its default value ($\epsilon=10^{-9}$). Such TP's were integrated for 5,000 years (saving an output every 5 years) to look for possible co-orbital outcomes during that time span.  This and other applications within this work of the \texttt{REBOUND} package refer to version 3.28.4.

\subsection{Launching Conditions}
\label{subsec:launching}
We consider four launching sites positioned along the Moon's equator -- the near-side, far-side, trailing-side, and leading-side. These are representative of the hemispheres of the Moon defined by its synchronous rotational motion with respect to its orbital motion around Earth. 

\begin{figure}
    \centering
    \includegraphics[scale=1.0]{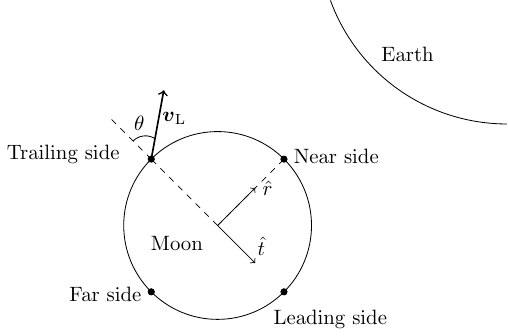}
    \caption{Launch conditions for lunar ejecta. Test particles are launched from four different equatorial sites and their speed is parameterized by $\theta$ and $v_{\mathrm{L}}$. The unit vector $\hat{r}$ defines the direction pointing towards Earth and $\hat{t}$ the transversal or leading direction. Adapted from \citep{Castro2023}. 
    }
    \label{fig:Diagram}
\end{figure}

At each launch location, the velocity of ejected particles is specified using two parameters, the angle $\theta$ 
with respect to the local normal (contained in the ecliptic plane) and the
launch speed $v_{L}$ (see Fig.~\ref{fig:Diagram}). The launch speed was varied
uniformly from 2.4 to 6.0 km/s in increments of 0.1 km/s and 100 particles
were launched with different directions (uniformly chosen between $\theta=-90^{\circ}$ and $\theta=90^{\circ}$) with the same speed. This combination of parameters yields 14,800  simultaneously launched particles.

\subsection{Choice of initial epoch}
\label{subsec:epochs}
In \citep{Castro2023}, TP's were launched at a specific epoch $t_{0}= \mathrm{J2452996}$ (22 December 2003) and the initial conditions for the planets and the Moon were retrieved from the JPL Horizons web-service. Here, we choose several distinct initial epochs in order to explore various initial conditions of the planets. 

\begin{figure}
    \centering
    \includegraphics[scale=0.8]{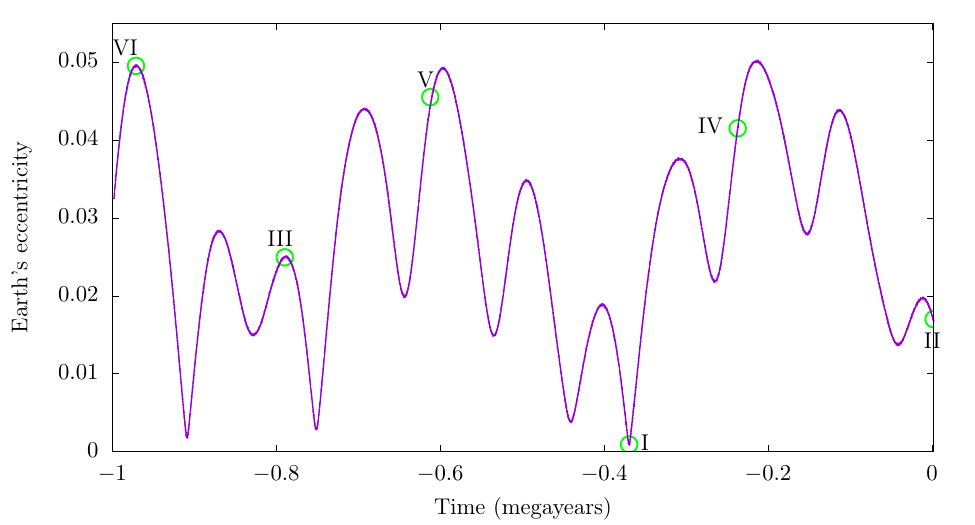}
    \caption{ Eccentricity of the Earth over a time span of 1 million years into the past (from December 22 2003) . The Earth-Moon barycenter was propagated under the influence of the Sun and the other seven planets in the Solar System using the \texttt{WHFast} integrator within the \texttt{REBOUND} package. 
    }
    \label{fig:Long}
\end{figure}

Since we want to test the effect of the Earth's eccentricity, we selected six epochs that are representative of the Earth's eccentricity range on secular timescales. To do this we evolved the Earth-Moon barycenter and the eight major planets for 1 megayear into the past (starting at $t_{0}$),  but now using the \texttt{WHFast} integrator within \texttt{REBOUND}, and recording the state of all bodies at intervals of 500 years. From the outputs, we chose six different epochs to sample the whole range of values of Earth's eccentricity. The evolution of the eccentricity and the location of the selected epochs can be seen in Fig.~\ref{fig:Long}, while the exact values and epochs are listed in Table~\ref{table:epochs}. Notice that the labels assigned to the epochs do not correspond to a chronological order, but correspond to increasing values of Earth's eccentricity.

\subsection{Initial conditions of the Moon}
\label{subsec:initial conditions}
 From this integration of the planets described in the previous section, we obtain initial conditions for the eight planets and the Earth-Moon barycenter. Initial conditions for the Earth and the Moon as individual bodies can
be derived if their relative position $\mathbf{r}$ and velocity $\mathbf{v}$ are known, according to the following equations: 
\begin{equation}
    \mathbf{r}_{\oplus} = \mathbf{r}_{\mathrm{EM}} + \frac{M_{\rightmoon}} {M_{\oplus}}  \bf{r};  \hspace{1cm} \mathbf{r}_{\rightmoon} = \mathrm{r}_{\mathrm{EM}} - \frac{M_{\oplus}} {M_{\rightmoon}} \bf{r}
\end{equation}

\begin{equation}
    \mathbf{v}_{\oplus} = \mathbf{r}_{\mathrm{EM}} + \frac{M_{\rightmoon}} {M_{\oplus}}  \bf{r};  \hspace{1cm} \mathbf{v}_{\rightmoon} = \mathrm{r}_{\mathrm{EM}} - \frac{M_{\oplus}} {M_{\rightmoon}} \bf{r}
\end{equation}

The relative position and velocity of the Earth and Moon are obtained from JPL Horizons web-service at the epoch $t_{0}$. The same relative position and velocity are adopted to calculate initial conditions of Earth and Moon for all the epochs in this study. %The previous procedure implies that the lunar phase is fixed, and permits us to focus solely on the effect of the varying eccentricity of Earth's orbit. 
The numerical experiment described at \ref{subsec:model} was performed for each of the selected epochs, resulting in a total of 88,800 simulated TP's.

\begin{center}
 \begin{table}[H]
\begin{tabular}{l|l|l}
Simulation Number    & $e_{Earth}$ & $t$, Epoch                                           \\ \hline
I   & 0.0092      & $t_{0}-371.5 \times 10^{3} \, \mathrm{years}$ \\ \hline
II  & 0.0167      & $t_{0}$                                       \\ \hline
III & 0.0250      & $t_{0}-791.5 \times 10^{3} \, \mathrm{years}$ \\ \hline
IV  & 0.0415      & $t_{0}-239.0 \times 10^{3} \, \mathrm{years}$ \\ \hline
V   & 0.0455      & $t_{0}-612.0 \times 10^{3} \, \mathrm{years}$ \\ \hline
VI  & 0.0495      & $t_{0}-973.0 \times 10^{3} \, \mathrm{years}$
\end{tabular}
\caption{Selected epochs along the Earth's eccentricity range. The epoch, $t$, is given relative to the initial epoch $t_{0}$ of the fiducial numerical simulation of the eight planet model of the Solar System (see main text for details). }
\label{table:epochs}
\end{table}   
\end{center}

\subsection{Classification of outcomes}
\label{classification}

\begin{figure}
    \centering
    \includegraphics[scale=0.75]{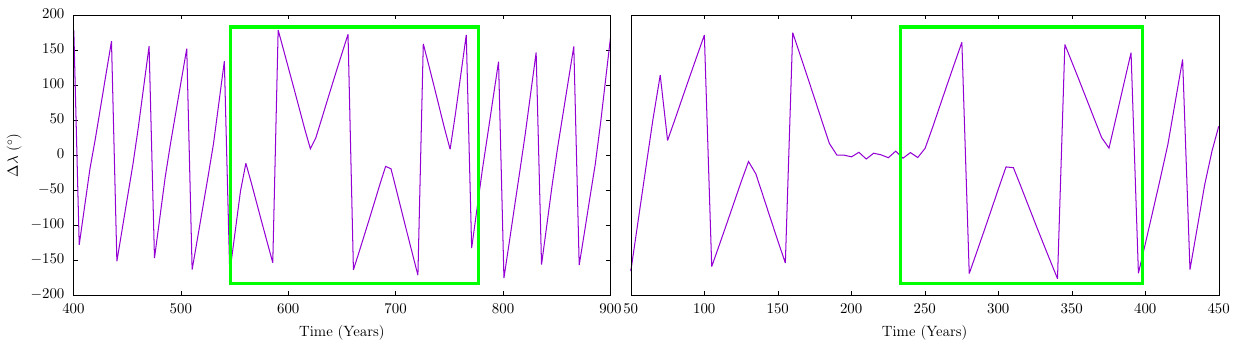}
    \caption{ Examples of two co-orbital identifications, (Left) for a HS orbit and (Right) for a HS-QS orbit. The \textit{green} windows represents the sufficient interval to make the identification in both cases.   
    }
    \label{fig:Co-orbitals}
\end{figure}

\begin{figure}
    \centering
    \includegraphics[scale=1.1]{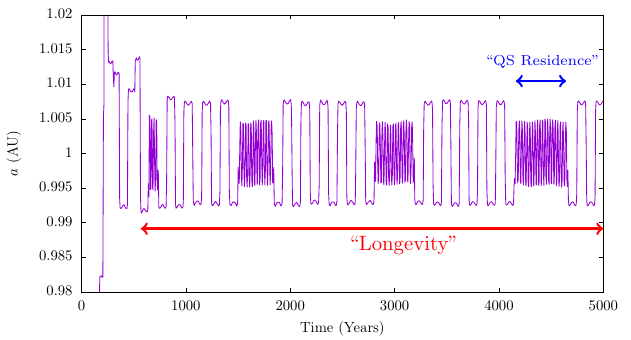}
    \caption{ Example of the evolution of the semi-major axis of a simulated lunar ejecta outcome displaying HS-QS transitions. % its Longevity and QS residence times are shown in \textit{red} and \textit{blue} respectivley. 
    The definition of the individual QS Residence time and the "Longevity" of the HS-QS state are indicated.}
    \label{fig:Definitions}
\end{figure}

%During  the 5,000 years time span the TP's were integrated, 
The time series of the semi-major axis $a$ and relative mean longitude $\Delta \lambda=\lambda-\lambda_{\mathrm{Earth}}$ were visually inspected to identify co-orbital behavior, in particular HS and HS-QS behavior. We recorded a HS outcome if we observed at least two oscillations in the time series of $\Delta \lambda$ around $180^{\circ}$; a HS-QS was recorded if we observed at least one oscillation of $\Delta \lambda$ around $0^{\circ}$ and one around $180^{\circ}$  (see Fig~\ref{fig:Co-orbitals} for examples of co-orbital identifications.) Although these criteria underestimate the actual number of co-orbiters, we found that more relaxed criteria (such as permitting only one oscillation to identify a HS) lead to unreliable counting since it is hard to visually spot only one period, particularly for rapidly oscillating time series with periods of few years. The unaccounted co-orbital outcomes are in any case, those which depart extremely rapidly, and are therefore, not of the greatest interest for this investigation. Since one of the motivations for this work
is the search for long-term Earth co-orbiters like Kamo'oalewa, we define two parameters to characterize the stability of HS-QS outcomes. The first parameter is the longevity of the co-orbital state, which is the total time spent in the combination of consecutive HS or QS states. The second parameter we use is the QS Residence time, which is the time spent in a single QS state. The definitions of Longevity and Residence time are illustrated in Fig.~\ref{fig:Definitions}. Investigations to understand the dynamical factors that influence these times have been made by \citep{yQdQ22a}.

\section{Numerical Results}
\label{sec:results}

%After the six numerical lunar ejecta experiments for each epoch, 
As expected, in all the simulations for all six initial epochs, the most common outcome for the TPs is transfer to general NEO-like orbits. 
%We observed multiple dynamical behaviors, including the identified co-orbital states, in all the simulations for all six initial epochs. 
Another possible fate for these particles is a collision with a major body. Transfer to a co-orbital state is not frequent, as we know from previous studies. Below, we first comment briefly on the colliders before describing in detail the co-orbital outcomes.

In these simulations of time span 5000 yr, we
 observed collisions only with the Earth and Moon, with around 8.3\% of all launched particles
ending in a collision. The more detailed statistics can be found in Table~\ref{table:collisions}, where we can also notice that, for almost all cases, ejecta launched from the lunar trailing
side produced the most particles which ended up in a collision, while particles launched from
the leading side had the fewest colliders. Most of the collisions with the Moon occur quite early in our simulations, with more than half of them colliding in the first year. Earth colliders are more widespread in time, with a median collision time
of around 900 years. We found no significant differences between the number of collisions for the different epochs. 

\begin{figure}
    \begin{flushleft}
    \includegraphics[scale=0.7]{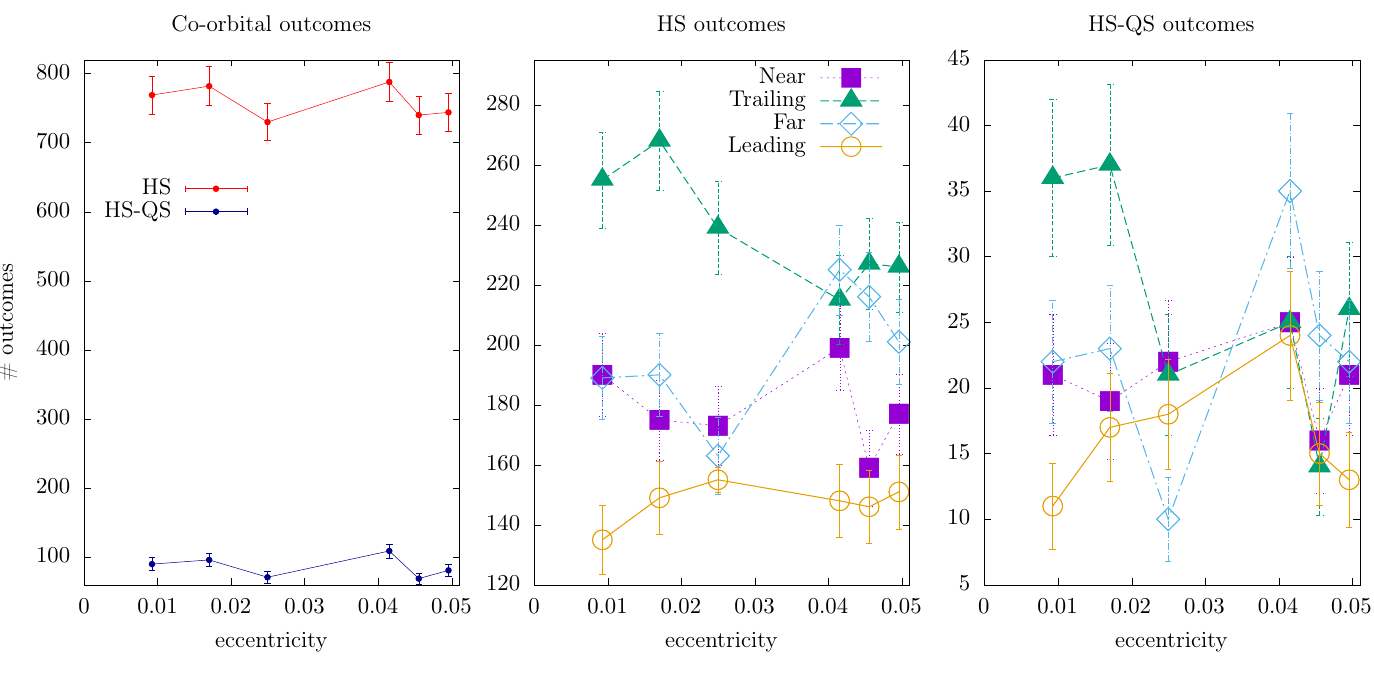}
    \caption{(Left) Number of co-orbital outcomes registered for each of the numerical experiments corresponding to the six representative launching epochs. (Center) Number of HS outcomes for each of the four launching sites, for each epoch. (Right) Number of HS-QS outcomes for each of the four launching sites, for each epoch. Error bars are based on Poisson statistics.
    }
    \label{fig:outcomes}
    \end{flushleft}

\end{figure}

The overall
production of co-orbiters is shown in Fig.~\ref{fig:outcomes}. In the \textit{left}
panel we can observe how the number of identified HS outcomes is always much larger than that
of HS-QS. The variation with respect to the eccentricity is not monotonic, with the production of 
HS co-orbiters ranging from  4.9\% to 5.3\%, and the production of HS-QS co-orbiters ranging from
0.5\% to 0.7\%. The overall production of co-orbitals thus varied between 5.4\% and 6.0\%. Panels (center) and (right) let us observe that the effect on each site is not the same,
having decreases in the production from one site, while the others may increase. 

\begin{figure}
    \centering
    \includegraphics[scale=0.8]{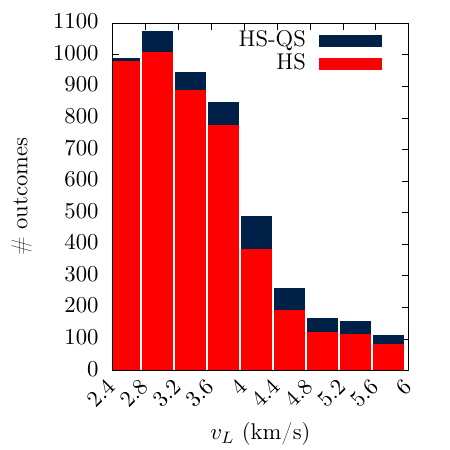}
    \caption{ Overall frequency of co-orbital outcomes (combined for all six epochs) as a function of launch speed. HS outcomes are shown in \textit{red} and
    HS-QS outcomes are shown in \textit{blue}. The frequency is increasing for the initial range and then decays rapidly at higher launch speeds. Similar trends are observed for the individual epochs (see the Appendix \ref{fig:histograms}). 
    }
    \label{fig:histogram}
\end{figure}

For each of the epochs, we can find how the frequency of the co-orbital outcomes varies as the
launch speed increases. This can be viewed in Fig.~\ref{fig:histogram}, showing,
 in general, a rapidly decaying trend with launch speed. Most of the co-orbiters are produced  at the lower speeds which is in agreement with the theoretical estimates made in \citep{Castro2023}.  The fact
 that the maximum of the production is not just above the lunar escape velocity is due to the number of
 colliding particles being maximum for such range.  
  For the larger speeds, the
frequency of co-orbiters becomes quite small, discouraging the exploration of larger values of the launch speed.  Both of these features appear in most of the individual histograms for each epoch (see Fig.~\ref{fig:histograms}).

\begin{figure}
    \centering
    \includegraphics[scale=0.65]{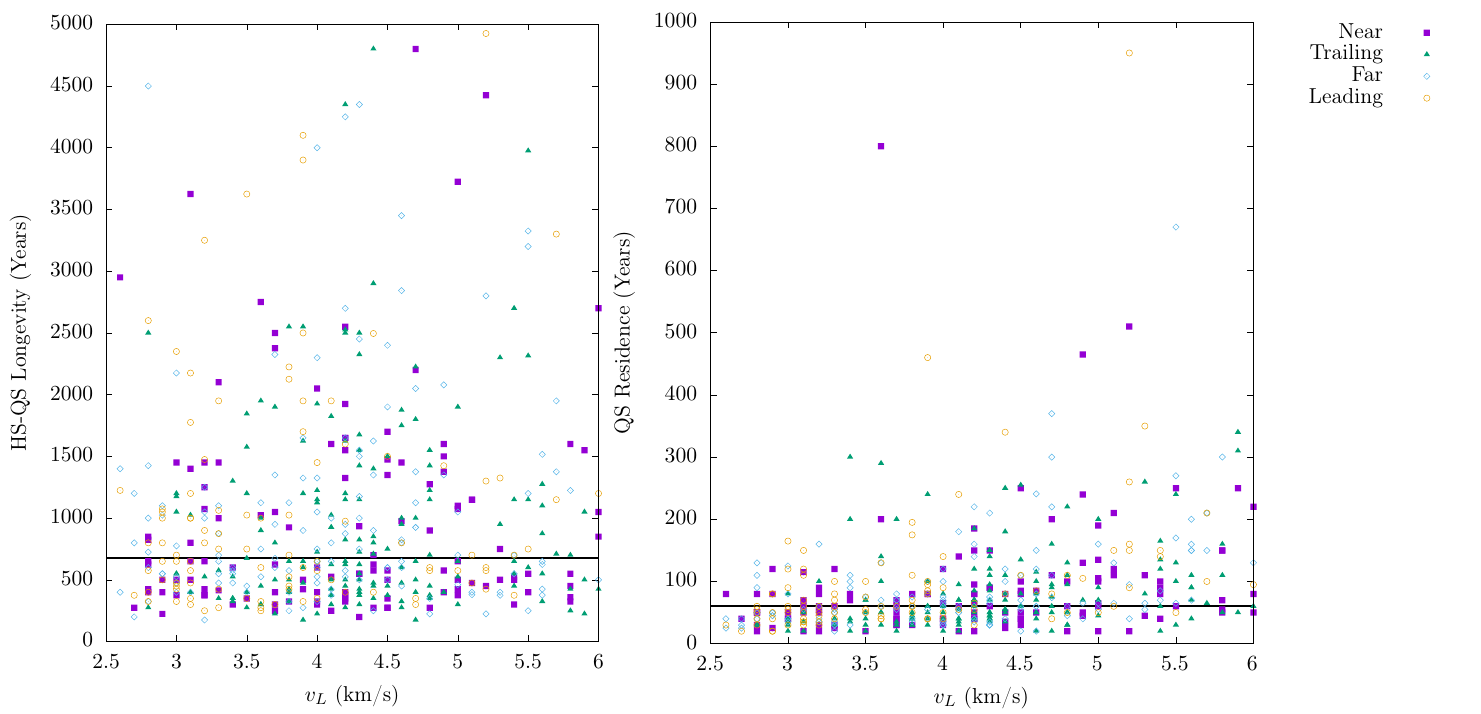}
\caption{  The longevities (left) and QS Residence times (right) (see the definitions in \ref{classification} and Fig.~\ref{fig:Definitions}) of all registered HS-QS outcomes are shown as functions of the launch speed; their median values are shown with a horizontal \textit{black} line.
    }
    \label{fig:Longevity}
\end{figure}

\begin{figure}
    \centering
    \includegraphics[scale=0.5]{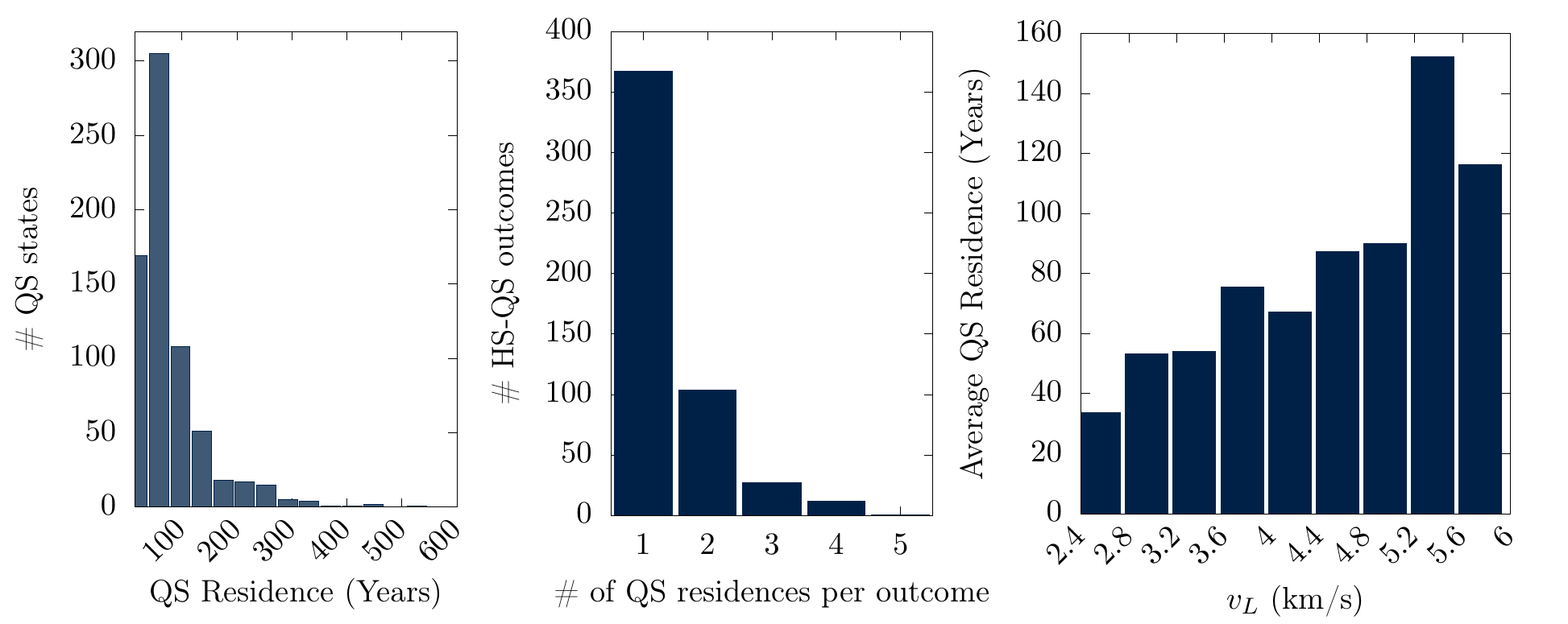}
    \caption{ (Left) Distribution of  the QS Residence times; (Center) Distribution of the frequency of QS states. (Right) Average QS residence times with respect to the launch speed. (Combined results from simulations for all six initial epochs and all four launch sites.) The uncommon, lmulti-century-long QS states like Kamo'oalewa's are favored to appear at the larger launch speeds.  
    }
    \label{fig:Longevity2}
\end{figure}

In terms of evaluating the lifetime of the HS-QS outcomes, we can plot the longevity and the residence as
a function of the launching speed Fig.~\ref{fig:Longevity}. In these plots we can observe how most of the co-orbiters
persist for less than 1,000 years, and most of the QS reside for less than 100 years. This can be further observed in Fig.~\ref{fig:Longevity2} (Left), where the frequency of the residence maximizes for the 50-100 years range to rapidly decay after this. Let us highlight however, that due to the criteria for HS-QS identification, and the resolution of time series (1 output every 5 years), the smallest residence detected was of 15 years. Fig.~\ref{fig:Longevity2} (Right) depicts the fact that the vast majority of the HS-QS outcomes (around 72\%) exhibit only one QS stage during the considered time span.  

\begin{figure}
    \centering
    \includegraphics[scale=1.0]{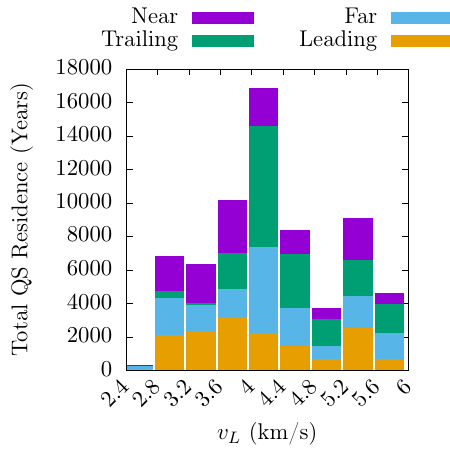}
    \caption{ Total QS residence times for as a function of the launch speed; the different colors indicate the contribution from each launch site. Longer QS residence times are favored for lunar ejecta launched from the trailing hemisphere at launch speeds near 4 km~s$^{-1}$.
    }
    \label{fig:total_times}
\end{figure}

Fig.~\ref{fig:Longevity} (Right) also exhibits skewness towards larger launch speeds. This can be further visualized when we consider the average residence time for each of the HS-QS outcomes, this is shown in  
Fig.~\ref{fig:Longevity2} (Right), indicating that although the number of QS outcomes decreases for the largest launching speeds, those fewer outcomes tend to have longer residence times. Notice however that the production of HS-QS outcomes is not monotonically decreasing (as is the HS production), but has a maximum between 4.0-4.4 km/s. Fig.~\ref{fig:total_times} depicts the total time the detected HS-QS outcomes remained in their QS states as the launch speed varies. This distribution has a maximum at the 4.0-4.4 km/s range, with the trailing side providing the largest contribution. This can also be observed for the overall larger launch speeds.

\section{Discussion} 
\label{sec:Discussion}

Based on the total number of outcomes shown in Fig~\ref{fig:outcomes} (Left), there is little sensitivity of HS and HS-QS outcomes production on Earth's eccentricity at the initial launch epoch. However, significant variations with epoch can be seen when we separate the statistics based on the launch site (see for example the differences between epochs III and IV in Fig~\ref{fig:outcomes} (Center) (Right)). These variations may be driven by the chaotic nature of the 
dynamics of this problem and the by the small sample of HS-QS outcomes. 
Despite these differences, the hierarchy among the launch sites production  remains consistent, with the Trailing side typically coming up as the dominant producer of HS outcomes, followed by the Far or Near side, and Leading side always resulting with the least amount. This hierarchy is particularly clear for the smaller values of Earth's eccentricity.

In contrast with the hundreds of HS outcomes, the frequency of HS-QS outcomes is much lower, with counts in just tens of HS-QS outcomes in the simulations from each of the four launch sites. The low statistics leads to less clear patterns in the outcomes. Amongst the low number statistics there is a slight hint that the ejecta from epochs of low values of Earth's eccentricity and launched from the lunar trailing side are more frequently represented in the HS-QS outcomes.

A supporting study by \citet{Jiao2024} concludes that the event leading to the formation of the lunar crater Giordano Bruno could have generated up to three \textit{Kamo'oalewa-like objects}. While differences in methodologies between the two studies result in non-identical and not directly comparable outcomes, they remain complementary. For instance, \citet{Jiao2024} reports a surviving fraction of approximately 84\% during the first 100 years of integration after particles are ejected. In contrast, our simulations indicate a higher survival rate of about 96\% after the same period. This discrepancy is understandable, as our study uses a uniform launch speed distribution and consequently includes relatively more particles at higher speeds whereas their study used a steep power law launch speed distribution favoring lower speeds; our higher frequency of larger launch speeds increases the likelihood of early escapes from the geocentric space, thus lower loss rate to collisions with Earth or Moon.

Due to their identification method, \citet{Jiao2024} did not detect short-lived co-orbiters. Our results, however, indicate that a significant fraction of the co-orbitals (specifically the HS outcomes)  have lifetimes of less than 100 years. This does not contradict the spirit of their investigation, as the primary interest lies in objects similar to Kamo'oalewa, which remain in their co-orbital HS-QS states for extended periods. Indeed, while many of the HS outcomes we registered were short-lived (less than 100 years), none of the HS-QS states in our results exhibited lifetimes shorter than 175 years.

In our simulations, about 6\% of launched particles reach a co-orbital state at some time during the 5,000 years simulation timespan. 
Jiao et al.'s study found that, 
within the first 10 million years after launch from the Giordano Bruno lunar impact crater, the percentage of particles residing in co-orbital state is at most 1\% (producing at most 3 total coorbital objects). 
These percentages are not directly comparable due to the different 
concepts they represent, however their lower percentage is not surprising because their study did not account for short-lived co-orbitals. Both studies reinforce the feasibility of the lunar origin of Kamo'oalewa, and reaffirm, from different perspectives, that the chances for such origin are small but sensible.

Our numerical experiments verify the existence of orbital pathways connecting particles launched from the lunar surface and reaching HS-QS orbits for the whole range of Earth's eccentricity at the epoch of launch, however, finding no strongly privileged epoch for the production of co-orbiters. On the one hand, this prevents finding a constraint on the age
of Kamo'oalewa based on a hypothetical privileged epoch (in terms of Earth's eccentricity value at the time of launch), but, on the other hand, it reinforces the claim that the trailing side of the Moon is the most suitable site to produce an object like this. This event
would still be extremely rare, given that the most long-lived co-orbiters produced by this
mechanism tend to appear for larger launching speeds, which are less likely to produce
co-orbitals at all, and would also be less frequent in a collision event (\citet{Jiao2024} finds that the velocity of ejected
fragments follows a power law of -4.0).

\section{Conclusions}

    \begin{itemize}
    \item The existence of orbital pathways connecting particles launched from the lunar surface and reaching HS-QS orbits is verified for the whole possible range of Earth's eccentricity at the epoch of launch. 
    \item The frequency of co-orbital outcomes decreases sharply as the launch speed increases, but long-lived QS's tend to be favored at larger launch speeds. 
    \item The overall production of co-orbiters is not very sensitive to Earth's eccentricity at the time of launch, and is 5.4\%--6.0\% over the full range of initial launch conditions that we investigated.
    \item  At low values of Earth's eccentricity the lunar Trailing hemisphere is the launch site with the largest production of co-orbiters. For the larger launch speeds it also has the largest contribution to the longer-lived QS outcomes.

  \end{itemize}

 \section{Data Availability}

Binary files with the initial conditions used for the simulations presented in this paper will be available at a publicly accessible permanent repository after acceptance. 

\section{Acknowledgements}
The results reported herein benefited from collaborations and/or information exchange within the program “Alien Earths” (supported by the National Aeronautics and Space Administration under Agreement No. 80NSSC21K0593) for NASA’s Nexus for Exoplanet System Science (NExSS) research coordination network sponsored by NASA’s Science Mission Directorate. 

\printcredits

\bibliographystyle{cas-model2-names}

\bibliography{cas-refs}

%\newpage

\appendix
\section{Appendix}
\label{sec:sample:appendix} 

\begin{figure}[h]
    \centering
    \includegraphics[scale=0.45]{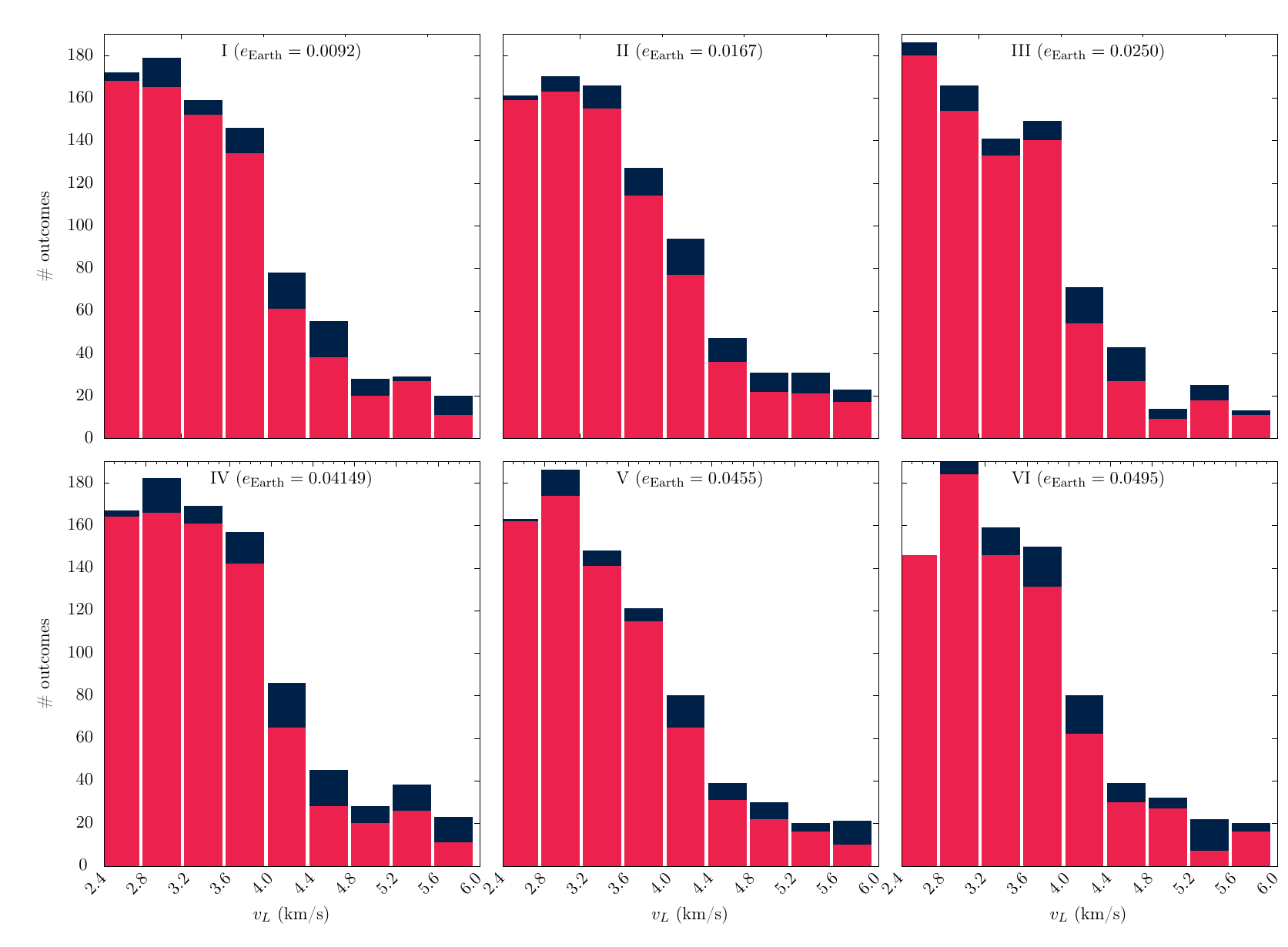}
    \caption{ Frequency of co-orbital outcomes for each of the selected epochs. HS outcomes appear in \textit{red} and
    HS-QS outcomes appear in \textit{blue}. Similar features are observed through all the cases.  
    }
    \label{fig:histograms}
\end{figure}

\begin{table}[h]
\begin{tabular}{l|lllll|lllll|}
\cline{2-11}
                          & \multicolumn{5}{l|}{Moon Colliders}                                                                                                                   & \multicolumn{5}{l|}{Earth Colliders}                                                                                                                   \\ \cline{2-11} 
                          & \multicolumn{1}{l|}{Near} & \multicolumn{1}{l|}{Trailing} & \multicolumn{1}{l|}{Far} & \multicolumn{1}{l|}{Leading} & Total                               & \multicolumn{1}{l|}{Near} & \multicolumn{1}{l|}{Trailing} & \multicolumn{1}{l|}{Far} & \multicolumn{1}{l|}{Leading} & Total                                \\ \hline
\multicolumn{1}{|l|}{I}   & \multicolumn{1}{l|}{35}   & \multicolumn{1}{l|}{30}       & \multicolumn{1}{l|}{36}  & \multicolumn{1}{l|}{11}      & 112 & \multicolumn{1}{l|}{276}  & \multicolumn{1}{l|}{391}      & \multicolumn{1}{l|}{295} & \multicolumn{1}{l|}{156}     & 1158 \\ \hline
\multicolumn{1}{|l|}{II}  & \multicolumn{1}{l|}{27}   & \multicolumn{1}{l|}{43}       & \multicolumn{1}{l|}{41}  & \multicolumn{1}{l|}{7}       & 118 & \multicolumn{1}{l|}{326}  & \multicolumn{1}{l|}{350}      & \multicolumn{1}{l|}{281} & \multicolumn{1}{l|}{129}     & 1086 \\ \hline
\multicolumn{1}{|l|}{III} & \multicolumn{1}{l|}{40}   & \multicolumn{1}{l|}{33}       & \multicolumn{1}{l|}{30}  & \multicolumn{1}{l|}{13}      & 116 & \multicolumn{1}{l|}{283}  & \multicolumn{1}{l|}{381}      & \multicolumn{1}{l|}{280} & \multicolumn{1}{l|}{123}     & 1067  \\ \hline
\multicolumn{1}{|l|}{IV}  & \multicolumn{1}{l|}{36}   & \multicolumn{1}{l|}{42}       & \multicolumn{1}{l|}{42}  & \multicolumn{1}{l|}{14}      & 134 & \multicolumn{1}{l|}{282}  & \multicolumn{1}{l|}{379}      & \multicolumn{1}{l|}{303} & \multicolumn{1}{l|}{150}     & 1114 \\ \hline
\multicolumn{1}{|l|}{V}   & \multicolumn{1}{l|}{30}   & \multicolumn{1}{l|}{37}       & \multicolumn{1}{l|}{29}  & \multicolumn{1}{l|}{9}       & 105 & \multicolumn{1}{l|}{307}  & \multicolumn{1}{l|}{378}      & \multicolumn{1}{l|}{276} & \multicolumn{1}{l|}{131}     & 1092 \\ \hline
\multicolumn{1}{|l|}{VI}  & \multicolumn{1}{l|}{28}   & \multicolumn{1}{l|}{35}       & \multicolumn{1}{l|}{28}  & \multicolumn{1}{l|}{11}      & 102 & \multicolumn{1}{l|}{314}  & \multicolumn{1}{l|}{382}      & \multicolumn{1}{l|}{299} & \multicolumn{1}{l|}{173}     & 1168 \\ \hline
\end{tabular}
\caption{Summary of lunar ejecta with colliding fate (only collisions with the Moon and the Earth were observed). For each epoch, the number of collisions is provided for each of the launching sites and the total is highlighted in \textit{red}. In general, trailing side has the largest number of both Moon and Earth colliders. }
\label{table:collisions}
\end{table}

%\end{linenumbers}
\end{document}